# INFLUENCE OF THE POSITION OF THE DOUBLE BOND ON THE AUTOIGNITION OF LINEAR ALKENES AT LOW TEMPERATURE

R. BOUNACEUR, V. WARTH, B. SIRJEAN, P.A. GLAUDE, R. FOURNET,

F. BATTIN-LECLERC*

*Département de Chimie-Physique des Réactions,*

*Nancy Université, CNRS,*

*1 rue Grandville, BP 20451, 54001 NANCY Cedex, France*

The influence of the position of the double bond on the autoignition of linear alkenes has been investigated by modeling the behavior of the 3 isomers of linear hexene and those of linear heptene. Low-temperature kinetic mechanisms for the oxidation of these six alkenes have been obtained after some improvements made to the system EXGAS, for the automatic generation of mechanisms, which had been previously adapted to model the oxidation of 1-pentene and 1-hexene. Quantum mechanical calculations have shown that *cis-trans* conformations should be taken into account and that isomerizations of alkenyl and alkenyl peroxy radicals involving a transition state including a double bond could be neglected. The new mechanisms have been validated using experimental data obtained in two rapid compression machines between 600 and 900 K with a good prediction of cool flame and ignition delay times. The model reproduces well the decreasing reactivity at low-temperature when going from 1- to 3-alkene. While the profiles of products are well reproduced for 1-hexene in a jet-stirred reactor above 780 K, more problems are encountered for the prediction of products in a rapid compression machine at 707 K, showing persisting problems in the understanding of the chemistry of the low-temperature oxidation of alkenes. Reaction rates analysis have been used to explain the difference of reactivity between the isomers of hexene.

---

* E-mail : Frederique.battin-Leclerc@ensic.inpl-nancy.fr ; Tel.: 33 3 83 17 51 25 , Fax : 33 3 83 37 81 20

## INTRODUCTION

Despite that the measurements of octane numbers show an important influence of the position of the double bond on the knocking properties of linear alkenes (RON is 76.4, 92.7 and 94, for 1-hexene, 2-hexene and 3-hexene, respectively [1]), there is to our knowledge no published model for linear alkenes other than 1-alkenes. Concerning 1-alkenes representative of the compounds present in gasoline, i.e. containing at least 5 atoms of carbon, the first modeling study has been published by Ribaucour et al. [2] to reproduce auto-ignition delay times of 1-pentene measured in a rapid compression machine (600-900 K) [3]. Another model for the oxidation of this species has been proposed by Mehl et al. [4] with validation using the same experimental data [3]. A high-temperature mechanism of the oxidation of 1-hexene has been published by Yahyaoui et al. [5] to reproduce their data obtained in a jet-stirred reactor between 750 and 1200 K and in a shock tube; this model underestimates the formation of products below 850 K. Using a version of EXGAS software extended to alkenes, Touchard et al. [6] have proposed mechanisms for the oxidation of 1-pentene and 1-hexene, which include 3385 reactions involving 837 species and 4526 reactions involving 1250 species, respectively. These mechanisms were validated using data measured for the auto-ignition of 1-pentene [3] and 1-hexene [7] in a rapid compression machine, showing well the difference of reactivity between the two compounds, and for the oxidation of 1-pentene in a flow tube (600-900 K) [8].

Concerning other linear alkenes, Vanhove et al. [7] have measured cool flame and auto-ignition delay times, as well as the formation of pre-ignition products, for the 3 linear isomers of hexene in a rapid compression machine (600-900 K) and Tanaka et al. [9] have measured the pressure profile during the combustion of the 3 linear isomers of heptene in a rapid compression machine at 827 K. These experimental studies show well that the reactivity of alkenes at low-temperature is considerably affected by the position of the double bond.

The purpose of this paper is to present the development of models for the oxidation of 1-, 2- and 3-alkenes and to present the validations made for the three isomers of hexene and heptene using the experimental results available in the literature [4, 7-8]. Quantum mechanical calculations have been used to gain a better insight in the isomerizations of alkenyl radicals. Flow rate analyses based on the obtained mechanisms have allowed the influence on the reactivity of the position of the double bond to be better understood.

## MODELS OF THE OXIDATION OF HEXENES AND HEPTENES

The present models (see the supplementary file in the case of the model for 3-heptene (834 species and 3032 reactions), the size of which is much smaller than that for 1-heptene (7178 reactions, 1620 species) and 2-heptene (8515 reactions, 1880 species) have been generated using a version of the EXGAS system in which specific improvements have been made compared to the version previously proposed to model the oxidation of 1-alkenes at low temperature [6]. The additional changes needed for this study will be described in more detail.

### *General features of the EXGAS system*

The EXGAS software has already been extensively described for alkanes [10-11], as well as for alkenes [6,12-14]; only a summary of its main features is then given here. The system provides reaction mechanisms made of three parts:

- A $C_0$-$C_2$ reaction base, including all the reactions involving radicals or molecules containing less than three carbon atoms [15].
- A comprehensive primary mechanism including all the reactions of the molecular reactants, the initial organic compounds and oxygen, and of the derived free radicals.

Figure 1 summarizes the different types of generic reactions, which are taken into account in the primary mechanism of the oxidation of alkenes, and the structure of the algorithm, which is used for the generation and which ensures the comprehensiveness of the mechanisms. Previous papers [6,10-14] describe in details these generic reactions and the related kinetic parameters.

**FIGURE 1**

- ♦ A lumped secondary mechanism, containing reactions consuming the molecular products of the primary mechanism which do not react in the reaction base.

Thermochemical data for molecules or radicals are automatically calculated and stored as 14 polynomial coefficients, according to the CHEMKIN II formalism [16]. These data are computed using the THERGAS software [17], based on group and bond additivity methods proposed by Benson [18].

*Specific improvements made in order to model the oxidation of 2-and 3-alkenes*

We present here only the two generic reactions which have been modified subsequent to our modelling of the oxidation of 1-pentene and 1-hexene [6],[14].

- ***Addition of OH radicals to the double bond***

Figure 2 illustrates the different pathways, which have been considered for the addition of OH radical to the double bond of 3-heptene. Pathways (1) and (2) correspond to the formation of an adduct, the reactivity of the obtained hydroxyalkyl radicals is assumed to be similar to that of alkyl radicals. Pathways (3) and (4) are addition/decompositions leading to the formation of aldehydes (acetaldehyde and pentanal in the case of 2-heptene) and alkyl radicals. Global reactions have to be written, because the decomposition of the adduct back to the reactant is faster than that toward products. The rate constants of these pathways have an important influence on the global reactivity and have been derived from the value ($k_{ref} = 1.4x10^{12}exp(-520/T)\ cm^3.mol^{-1}.s^{-1}$) proposed by Tsang for the addition of OH radical to propene [19] with adjusted values (factors no higher than 10) in order to get satisfactory simulations. This necessary adjustment hinders the predictive abilities of the model generated by EXGAS for other alkenes.

**FIGURE 2**

- ***Isomerizations involving a transition state including a double bond***

The isomerizations of alkenyl and alkenyl peroxy radicals involving a transition state including a double bond (as illustrated in figure 3 in the case of a peroxy radicals which is produced from 2-heptene) were considered in our previous work. The isomerizations of alkenyl peroxy radicals involving a transition state including a double bond have been proven to cause a large overprediction of the formation of cyclic ethers with an unsaturation located inside the cycle. Due to the following theoretical considerations, both types of isomerizations have been neglected in the present work. The neglecting of these reactions explains why far fewer reactions are considered in the case of 3-heptene than for 1- and 2-heptene.

Alkenyl or alkenyl peroxy radicals need necessarily a *cis* conformation in order to form an unsaturated cyclic transition state. A *trans* conformation does not allow the distance between the abstractable H-atom and the radical center to be compatible with the occurrence of the sought reaction. If we consider that initial molecules are all in *trans* conformation, which is supposed to be the most stable conformation, the standard enthalpy activation is about 60 kcal.mol$^{-1}$ for *trans*-*cis* conversion according to the value measured by Masson et al. [20] for 2-butene and makes the occurrence of this last reaction unlikely at low temperature. A second possibility is the *trans*-*cis* conversion from a stabilized free radical as illustrated in figure 4 for 2-penten-1-yl. In this last case, the electronic delocalization can make possible the rotation around the carbon-carbon bond involved in the initial π bond, with a lower energy barrier. Quantum calculations were performed at the CBS-QB3 level of theory using Gaussian03 [21] in order to estimate this activation energy with a given accuracy of $\pm$ 1 kcal/mol. The standard enthalpy activation has been obtained from electronic energy added with the zero point energy and thermal correction to enthalpy at 298.15 K. Intrinsic Reaction Coordinate (IRC) calculations have been systematically performed at the B3LYP/6-31G(d) level of theory on transition states (TSs), to ensure that they are correctly connected to the desired reactants and products. As shown in figure 4, the energy barrier is equal to 15.3 kcal.mol$^{-1}$ what makes inter conversion difficult again at low temperature compared to the addition of the *trans* radical to oxygen molecules. It is worth noting that the *trans*-*cis* conversion involves a removal of the resonance in the TS. The bond lengths between carbon atoms 6-4 and 4-1 are very close ($d_{6\text{-}4}$ = 1.386 Å and $d_{4\text{-}1}$=1.383 Å) in the reactant and are characteristic of the delocalisation of the π bond (figure 4) while in the TS the length between carbon atoms 6-4 (d = 1.472 Å ) is greater than the length between carbon atoms 4-1 (d = 1.332 Å) which is close to the length of a regular double bond.

Another point concerns the isomerizations of stabilized *cis* alkenyl free radicals. Isomerization of *cis* pent-2-en-1-yl (figure 4) has been studied at the CBS-QB3 level of theory and an activation energy ($E_a$) of 36.7 kcal.mol$^{-1}$ has been obtained for the 1-4 hydrogen transfer. This last value can be compared with an empirical estimation of *E*a which involves the ring strain energy of the cyclic structure in the TS ($E_{RSE}$) added to the abstraction energy of a secondary H-atom by a stabilized free radical ($E_{abs}$). In this example, $E_{RSE}$ = 5.9 kcal.mol$^{-1}$ and $E_{abs}$ = 15.5 kcal.mol$^{-1}$ [6] leading to an activation energy $E_a$ of 21.4 kcal.mol$^{-1}$. This value is lower of 15.3 kcal.mol$^{-1}$ compared to the CBS-QB3 calculation. The obtained difference of activation energies is due to the necessary rotation of the $CH_2$ group in the TS (carbon 1 in figure 4) so that it can be facing the abstracted hydrogen atom (hydrogen 9 in the TS of figure 4). As in the case of the *trans-cis* conversion, the rotation of the $CH_2$ group leads to a removal of the resonance in the TS corresponding to an increase of the activation energy of 15.3 kcal.mol$^{-1}$.

**FIGURE 3**

**FIGURE 4**

## COMPARISON BETWEEN SIMULATIONS AND EXPERIMENTS

Simulations have been performed using the PSR and SENKIN softwares of CHEMKIN II [16] without considering heat transfer. In figures 5, 7, 8, and 9, symbols correspond to experimental results and lines to simulations.

*Modeling of the oxidation of the three isomers of hexene in a rapid compression machine*

Vanhove et al. [7] have measured cool flame and auto-ignition delay times in a rapid compression for hexene/$O_2$/Ar/$N_2$/$CO_2$ mixtures for temperatures ($T_c$) after compression ranging from 615 to 850 K. $T_c$ is calculated at the end of the compression based on a core gas model [7] and can be considered as the maximum temperature reached in the combustion chamber. Four initial pressures have been tested leading to pressures ($P_c$) after compression ranging from 6 to 8.4 bar, from 7.7 to 9.6 bar, from 8 to 11 bar and from 10 to 12 bar, respectively. For all studied pressures ranges, the agreement obtained between our simulations and these experimental results is satisfactory both for cool flame (cool flame was neither experimentally observed nor simulated for 3-hexene) and ignition delay times, as illustrated in Figure 5 in the case of the 8-11 bar range. Simulations reproduce well the difference of reactivity between the three isomers; while 2-hexene has only a slightly lower octane number than 3-hexene, its reactivity is much more similar to that of 1-hexene than to that of 3-hexene.

**FIGURE 5**

As shown in Figure 6, the model reproduces correctly the major trends of the distribution of products for 1-hexene, especially taking into account that some quantitative discrepancies may be explained by the gradients of temperature which are observed inside the combustion chamber of a rapid compression machine. However, the model predicts the formation of cyclic ethers bearing an alcohol function, which are not experimentally observed, and the agreement deteriorates for other isomers. While the same selectivity of unsaturated cyclic ethers is predicted whatever the isomer, experimental results show a large reduction of their formation from 2- and 3-hexene. While the selectivity of the major saturated aldehydes (acetaldehyde and pentanal) are well reproduced for 1-hexene, the formation of butanal is considerably underestimated from 2-hexene, as well as propanal from 3-hexene.

**FIGURE 6**

*Modeling of the oxidation of 1-hexene in a jet-stirred reactor*

Yahyaoui et al. [5] have studied the oxidation of 1-hexene in a jet stirred reactor at 10 bar for temperature from 780 to 1100 K. Figure 7 shows that the computed profiles are in good agreement with the experimental ones. The formation of the major products (see figure 7b in logarithmic scale) is well predicted, even at the lowest temperatures. The fact that the formation of products between 780 and 850 K is better predicted than with the high temperature mechanism of Yahyaoui et al. [5] shows that the low temperature reactions are still of importance in this temperature range.

**FIGURE 7**

*Modeling of the oxidation of the three isomers of heptene in a rapid compression machine*

Tanaka et al. [9] have measured the pressure rise as a function of time for the three isomers of linear heptene in a rapid compression machine for $T_c$ equal to 827 K and $P_c$ to 41.6 bar. Figure 8 shows that our models can well predict the occurrence of cool flame for 1- and 2-heptene and the fact that such a phenomenon is not observed for 3-heptene. The prediction of the times of pressure rise related to cool flame or to ignition is also satisfactory. For these compounds more reactive than hexenes, simulation reflects well also the difference of reactivity between the three isomers in connection with their difference of octane number, i.e. RON is 54.5 for 1-heptene, 73.4 for 2-heptene and 89.8 for 3-heptene [1] .

**FIGURE 8**

## DISCUSSION

Table 1 presents the main rates of reactant consumption for the three linear isomers of hexene for an initial temperature of 752 K. The additions to oxygen molecules of the hydroxyalkyl or alkyl radicals formed by the different channels of OH-additions or of the alkenyl radicals obtained by H-abstractions are a source of hydroperoxide species, which are branching agents with a well-known promoting influence on the reaction rate. On the contrary, at this temperature, the reactions of alkenyl radicals with oxygen molecules to produce hexadienes and the very unreactive $HO_2$ radicals, as well as the combination of resonance stabilised alkenyl radicals with $HO_2$ radicals are a sink of radicals and have then an inhibiting effect. Table 1 shows that the total rate of reactant consumption by reactions with an inhibiting influence is 34% for 1-hexene, 49% for 2 hexene and 66% for 3-hexene. This increase of the rates of reactions with an inhibiting influence with the position of the double bond (1-, 2- or 3-) is certainly important to explain the decrease of reactivity from 2- and 3-hexene shown in figure 5. It is partly due to the fact that the formation of resonance stabilised alkenyl radicals from 2- and 3-hexene involves the abstraction of four different H-atoms (two primary and two secondary H-atoms from 2-hexene, four secondary ones from 3-hexene) while it can be obtained from the abstraction of only two secondary H-atoms in the case of 1-hexene. Above 800 K, the combination of $HO_2$ and resonance stabilised alkenyl radicals acts more like a propagation step as the obtained hydroperoxide molecule can rapidly decompose to give two radicals.

**TABLE 1**

Figure 9 presents a sensitivity analysis on two types of reactions which are of great importance to model the ignition characteristics of alkenes. It shows that, in the case of 3-hexene, the fact to consider the isomerizations involving a TS including a double bond, without distinguishing between *cis* and *trans* conformations and without taking into account the loss of resonance in the TS for allylic radicals isomerizations, as we did in our previous study of 1-alkenes [6], leads to a considerable underprediction (up to a factor 3) of the delay times. This is also noticeable for 2-hexene, but much less so for 1-hexene, and illustrates that the generic reactions proposed for the modelling of alkane oxidation cannot be applied to alkenes in a straightforward way. Further theoretical calculations for the estimation of other kinetic parameters of this mechanism, such as the formation of unsaturated cyclic ethers, would certainly be also valuable.

**FIGURE 9**

As our model underestimates the formation of aldehydes which can be formed by addition/decomposition of OH radicals to 2- and 3-hexene (i.e. butanal and propanal), figure 9 displays also the results of a simulation with the rate constant of that rate constant multiplied by a factor 10. A large underprediction of the delay times is then observed in the case of 3-hexene. Further experimental and theoretical investigations concerning the addition of OH radicals to alkenes and more globally the formation of oxygenated species during the oxidation of these unsaturated compounds are certainly needed.

## CONCLUSION

The new mechanisms, automatically generated using the system EXGAS, for the oxidation of linear isomers of hexane and heptenes have been used to satisfactorily model experimental results obtained in rapid compression machines [7,9] and in a jet-stirred reactor [5]. The mechanisms presented here are, to our knowledge, the first attempt to model the oxidation of 2-hexene, 3-hexene and linear isomers of heptene at low temperature. This study underlines the complexity of the modelling of the oxidation of alkenes and the need of further investigations concerning the involved types of elementary steps and their rate coefficients.

## ACKNOWLEDGEMENTS

This work has been supported by Institut Français du Pétrole.

Table I: Main channels of reactant consumption ($T_c$= 752 K, $P_c$ = 8.6 bar, $\phi$ = 1, dilution in air, 3% conversion of reactant) deduced from the flow rate analysis. Only the types of reaction for which the contribution to reactant rate consumption is above 10% for at least a compound are shown. The types of reaction in italics are those with an inhibiting effect.

| **Reactant** | **Type of reaction** | **Rate of reactant consumption (%)** |
|---|---|---|
| 1-hexene | OH addition | 41 |
| | H-abstraction/ $O_2$ addition | 17 |
| | *H-abstraction/ oxidation*[a] | *3* |
| | *H-abstraction/ combination*[b] | *31* |
| 2-hexene | OH addition | 22 |
| | H-abstraction/ $O_2$ addition | 19 |
| | *H-abstraction/ oxidation*[a] | *5* |
| | *H-abstraction/ combination*[b] | *44* |
| 3-hexene | OH addition | 6 |
| | H-abstraction/ $O_2$ addition | 23 |
| | *H-abstraction/ oxidation*[a] | *10* |
| | *H-abstraction/ combination*[b] | *56* |

[a]: Formation of diene and $HO_2$ radicals,
[b]: With $HO_2$ radicals.

**FIGURE CAPTIONS**

Figure 1: Algorithm of generation for the primary mechanism of the oxidation of alkenes. The reactions important at low temperature are shown in bold.

Figure 2: Different pathways for the addition of OH radical to 2-heptene. The rate constants of pathways (1) and (2) are equal to $k_{ref}$ for the formation of a 1-hydroxyalkyl radical, to $k_{ref}/1.4$ for that of a 2-hydroxyalkyl radical from a 1-alkene, to $k_{ref}/3$ for that of a 2-hydroxyalkyl radical from a 2-alkene and to $k_{ref}/5$ in all other cases. The rate constants of pathways (3) and (4) depend on the studied compound: $k_{ref}$ for 1-hexene, $k_{ref}/10$ for other isomer of hexene, $k_{ref}$x3 for 1- and 2-heptene, $k_{ref}/1.75$ for 3-heptene.

Figure 3: Example of isomerization involving a transition state including a double bond in the case of hept-2-enyl-1-peroxy radicals.

Figure 4: Standard enthalpy variation diagram for the *trans-cis* conversion of pent-2-en-1-yl radical, the isomerization of cis pent-2-en-1-yl radical in cis pent-2-en-4-yl radical, and the trans-cis conversion of this last radical. Standard enthalpy variations are given at 298.15 K in kcal mol-1. The molecular properties were determined at the CBS-QB3 level of theory.

Figure 5: Cool flame and ignition delay times for the three isomers of linear hexene in a rapid compression machine ($P_c$ between 8 and 11 bar, stoichiometric mixtures in air [7]).

Figure 6: Comparison between the predicted (black bars) and the experimental (white bars) (Table 1 [7]) selectivities of products during the pre-ignition period of 1-hexene under the conditions of figure ($T_c$= 707 K, $P_c$ = 7 bar, $\phi$ = 1, dilution in air).

Figure 7: Profiles of (a) reactants and (b) major products vs. temperature during the oxidation of 1-hexene in a jet-stirred reactor (pressure of 10 bar, stoichiometric mixture with 0.1% of hydrocarbon diluted in argon, residence time of 0.5 s [5]).

Figure 8: Pressure change ($\Delta P$) curves for the three isomers of linear heptene in a rapid

compression machine ($T_c$ = 627 K, $P_c$ = 41.6 bar, $\phi$ = 0.4, dilution in air [9]).

Figure 9: Sensitivity analysis for ignition delay times of 3-hexene under the conditions of figure 4. The broken line is computed with the initial mechanism, the full line when the isomerizations involving a transition state including a double bond are also considered, and the dotted line when the rate constants of OH radical addition/decomposition pathways are multiplied by 10.

**LIST OF SUPPLEMENTAL MATERIAL :**

1 supplemental file: mechanism-bounaceur et al.-2007

**CAPTION FOR SUPPLEMENTAL MATERIAL :**

Mechanism (in CHEMKIN format) for the low-temperature oxidation of 3-heptene

Figure 1

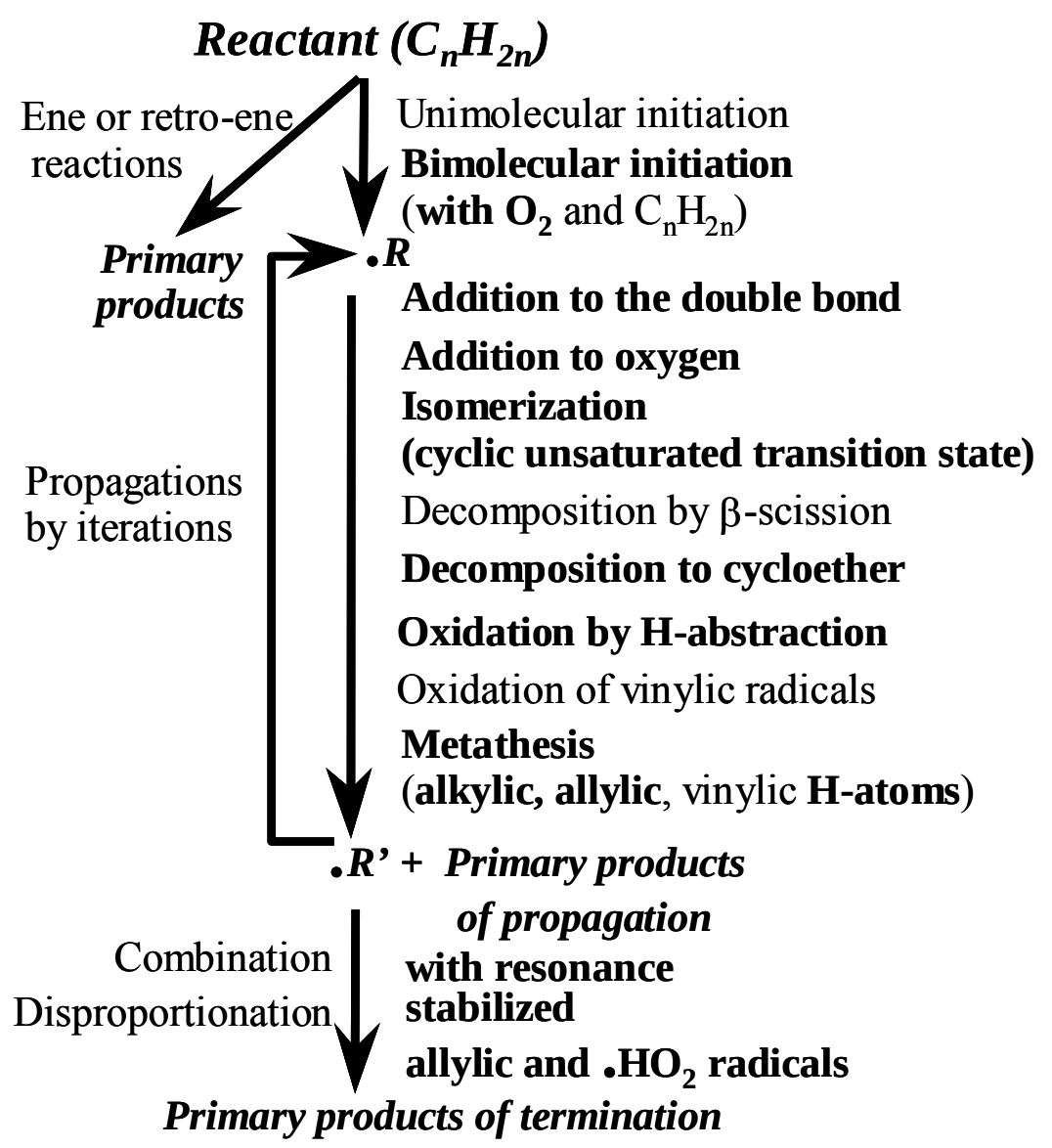

Reactant ($C_nH_{2n}$)
Ene or retro-ene reactions
Unimolecular initiation
Bimolecular initiation
(with $O_2$ and $C_nH_{2n}$)
Primary products
•R
Addition to the double bond
Addition to oxygen
Isomerization
(cyclic unsaturated transition state)
Propagations by iterations
Decomposition by β-scission
Decomposition to cycloether
Oxidation by H-abstraction
Oxidation of vinylic radicals
Metathesis
(alkylic, allylic, vinylic H-atoms)
•R' + Primary products of propagation
Combination
Disproportionation
with resonance stabilized allylic and $•HO_2$ radicals
Primary products of termination

Figure 2

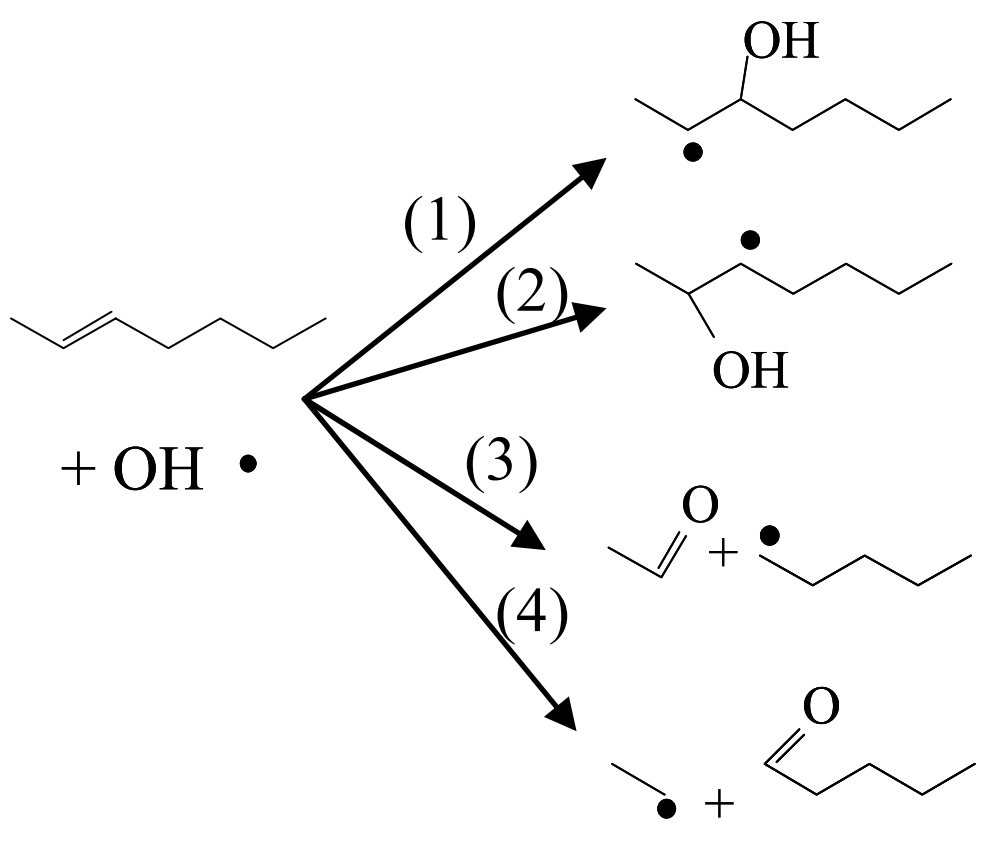

Figure 3

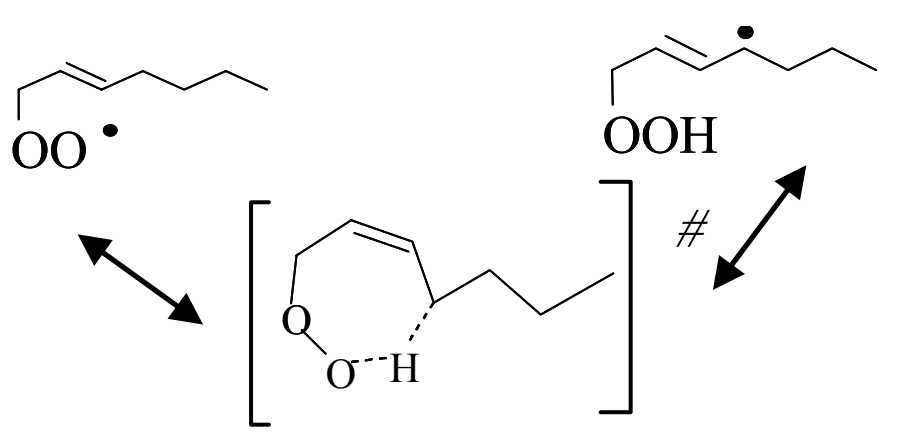

Figure 4

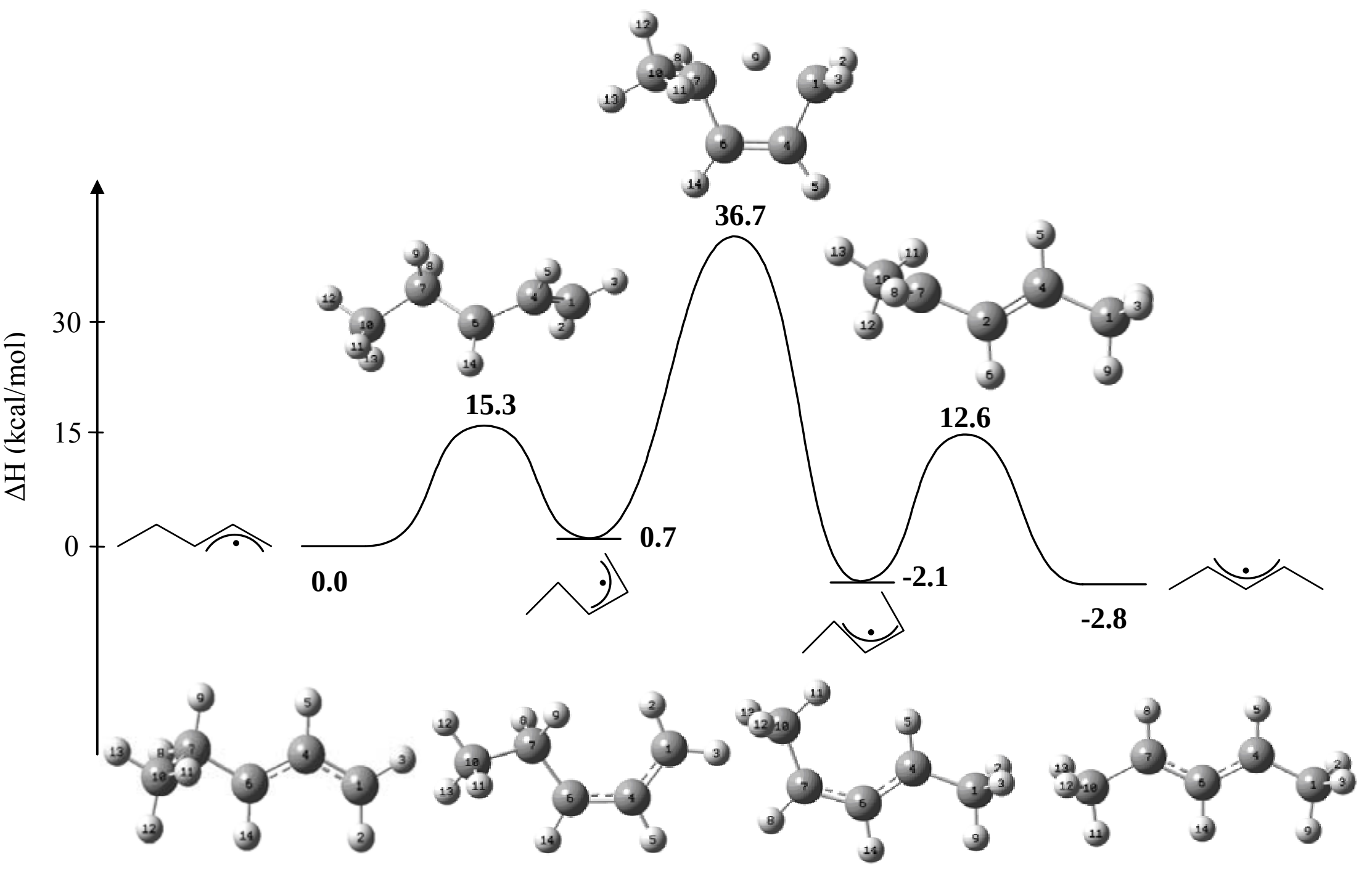

Figure 5

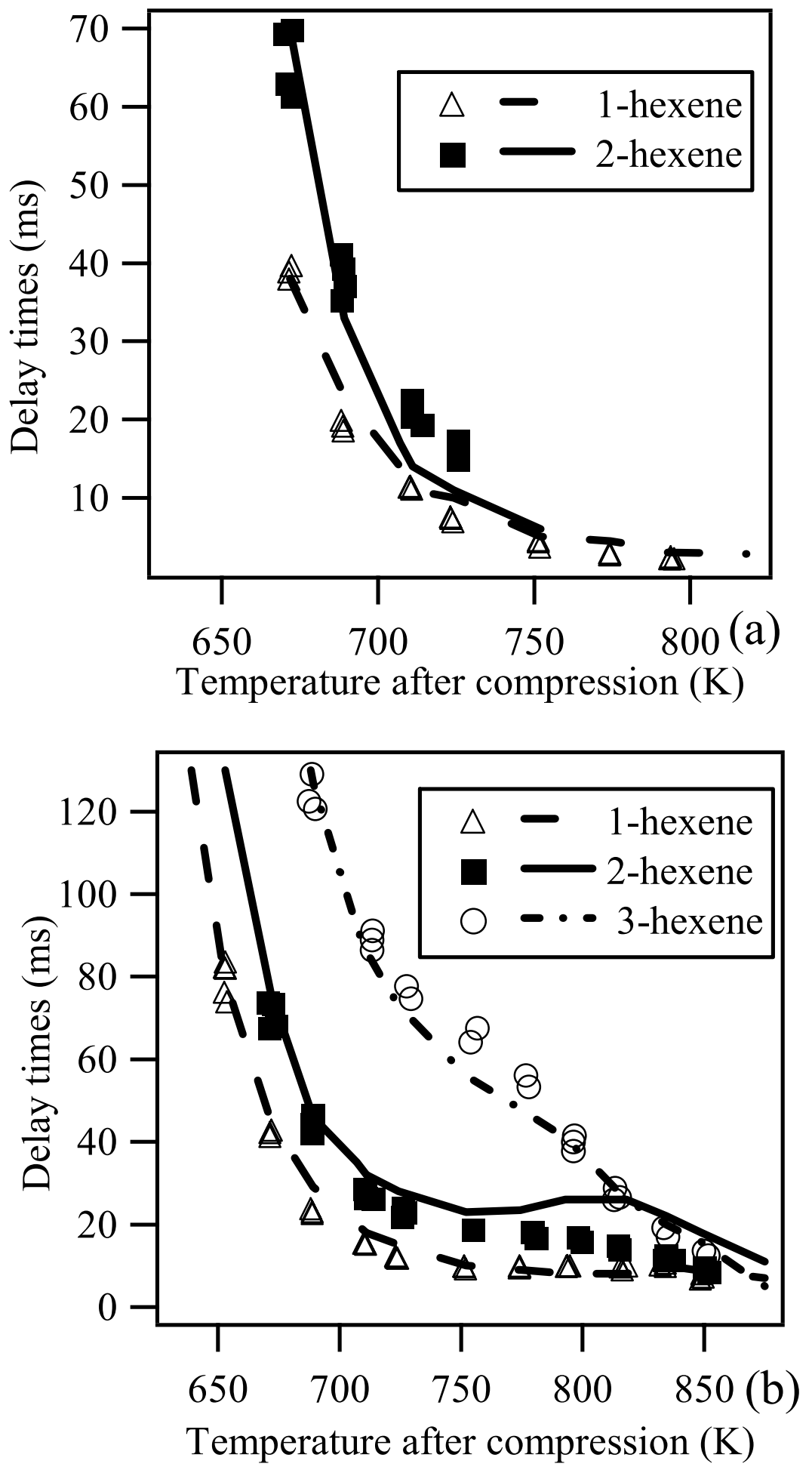

Figure 6

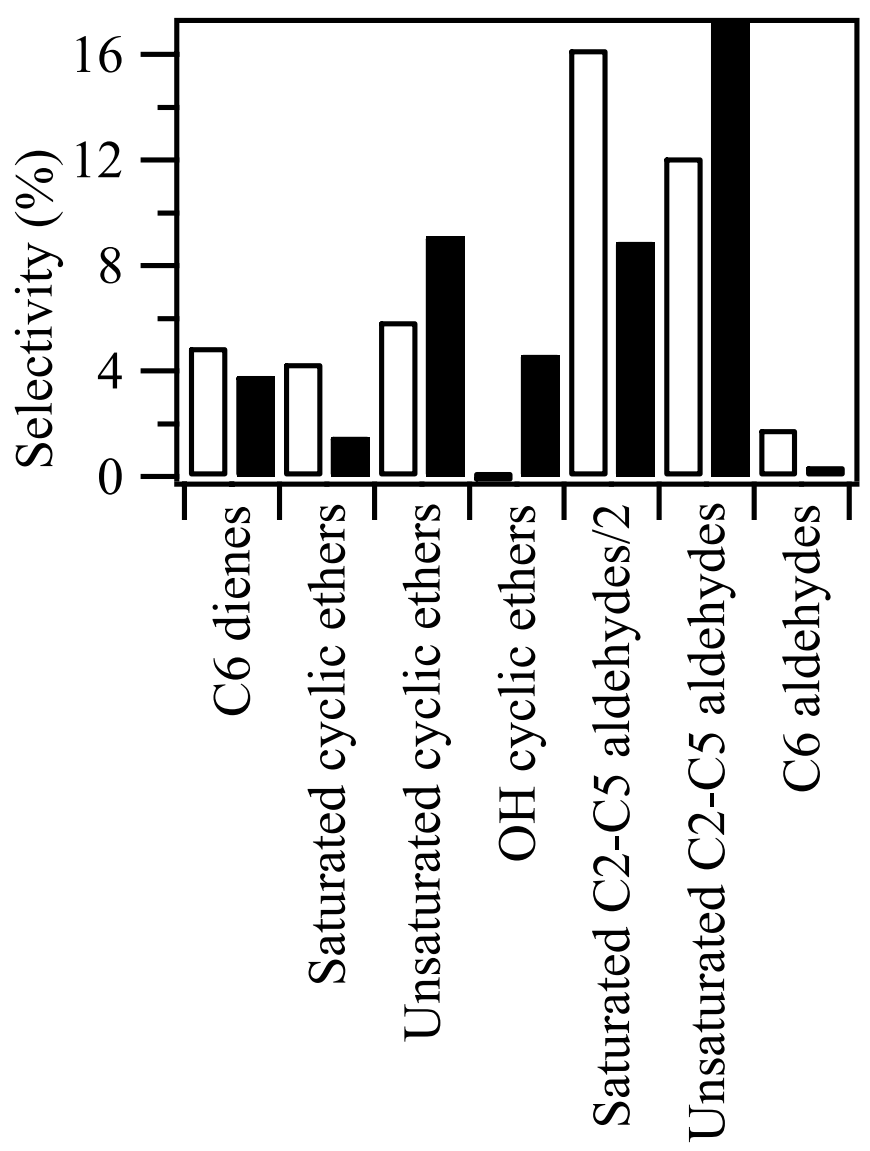

Figure 7

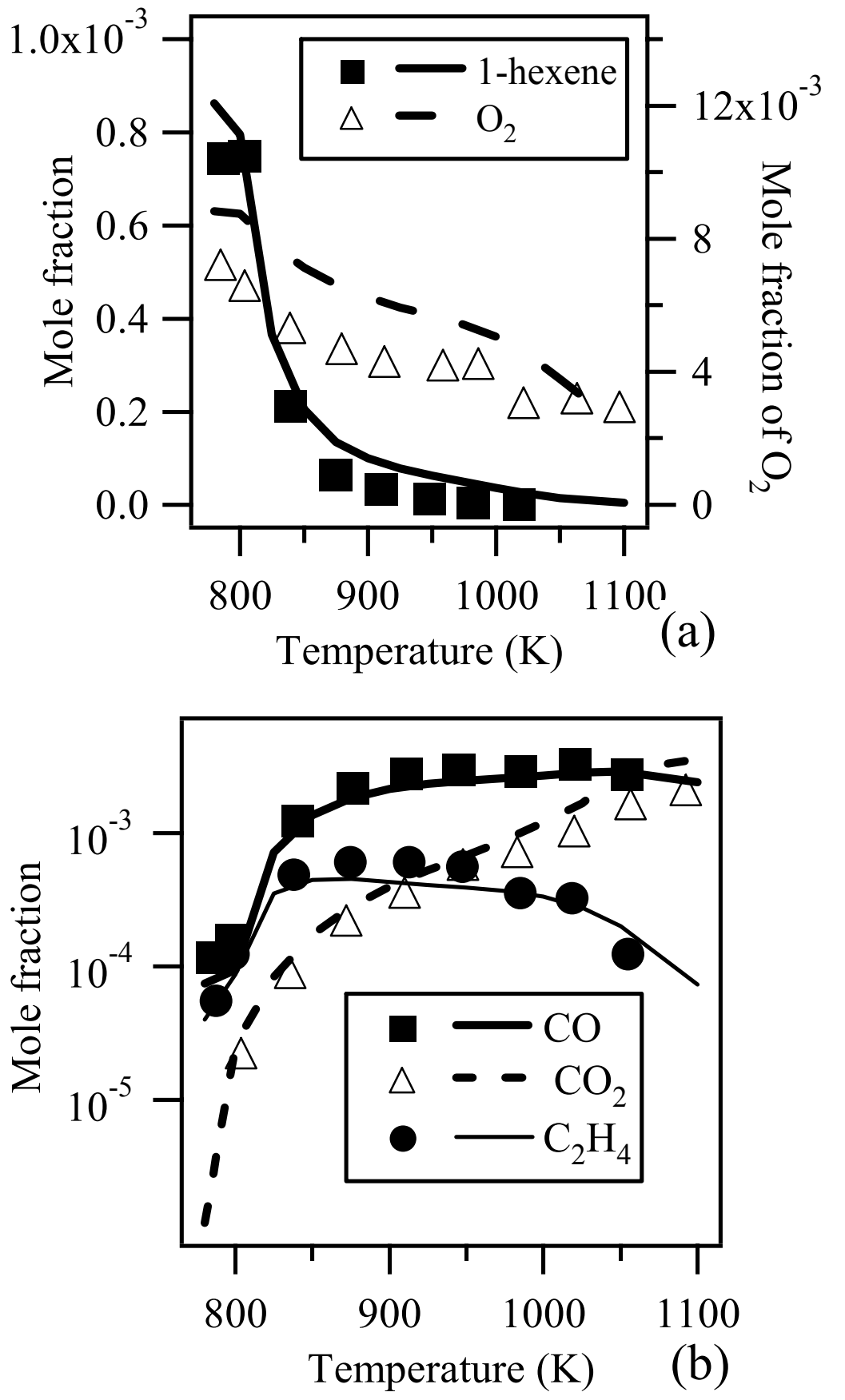

Figure 8

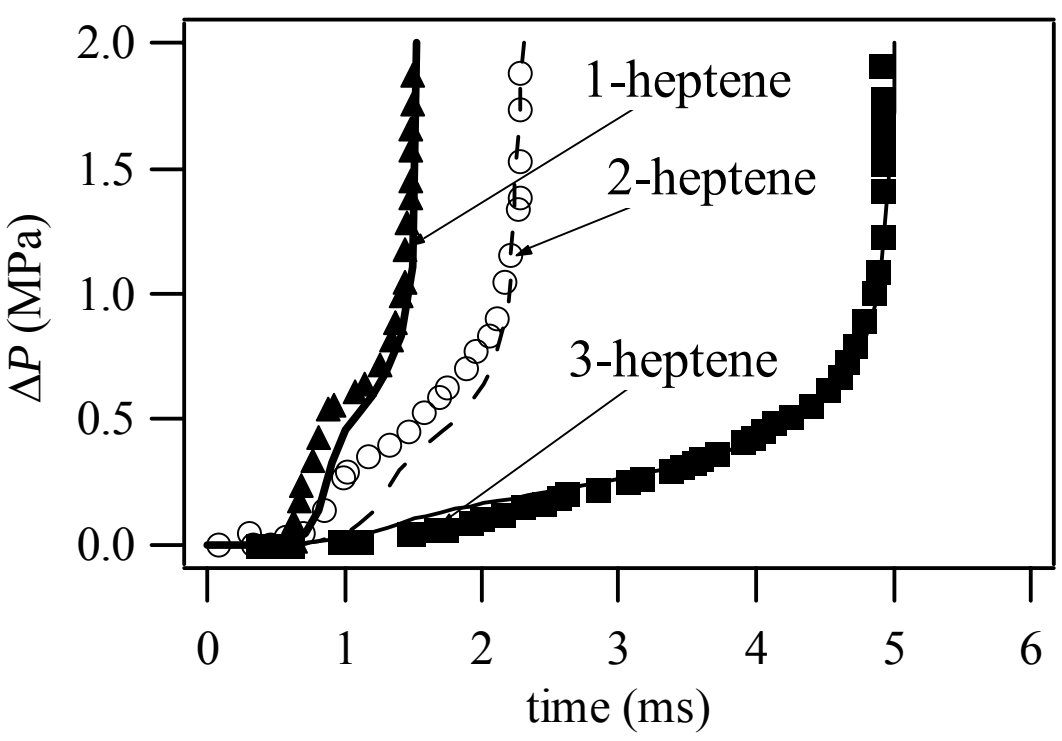

Figure 9

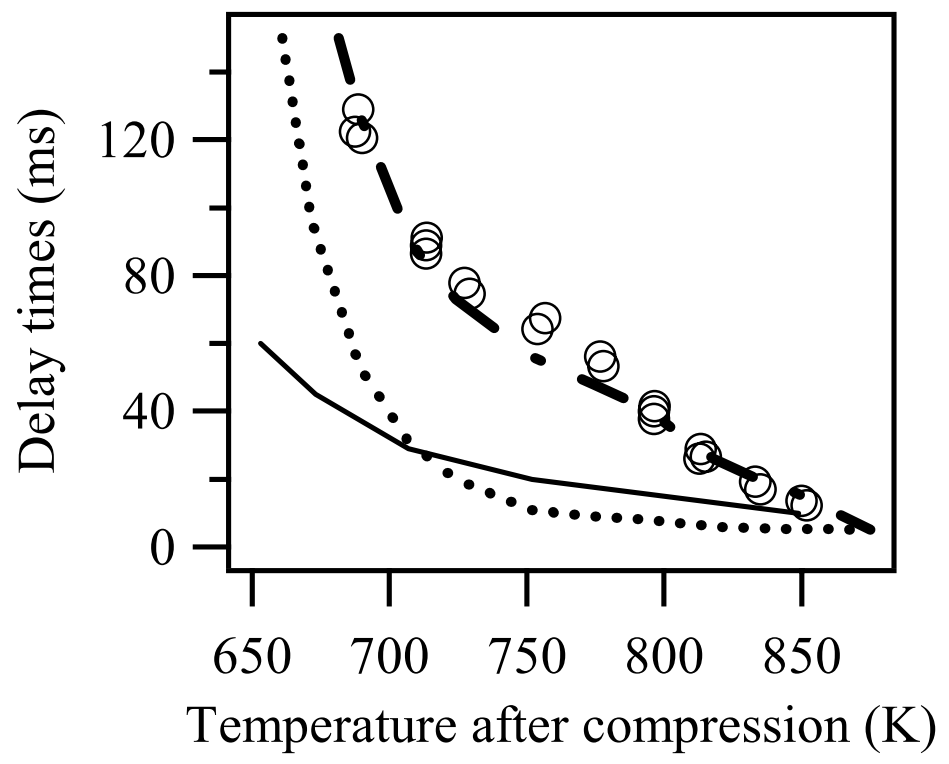